\documentclass[nobibnotes, aps, prl]{revtex4}

\usepackage{stmaryrd}
\usepackage{amssymb}
\usepackage{amsmath}
\usepackage{dsfont}
\usepackage{mathrsfs}
\usepackage{epsfig}

\newcommand{\gp}{geometric phase\ }
\newcommand{\be}{\begin{eqnarray}}
\newcommand{\ee}{\end{eqnarray}}

\begin{document}
\large

\title{Experimental Measurement of Mixed State Geometric Phase by Quantum Interferometry using NMR}
\author{Arindam Ghosh}
\author{Anil Kumar}
\altaffiliation{DAE-BRNS Senior Scientist.}
\email{anilnmr@physics.iisc.ernet.in}
\affiliation{NMR Quantum Computation and Quantum Information Group\\ Department of Physics and NMR Research Centre, Indian Institute of Science,
Bangalore-560012, India}

\begin{abstract}
Geometric phase has been proposed as one of the promising methodologies to perform fault tolerant quantum computations.
However, since decoherence plays a crucial role in such studies, understanding of
mixed state \gp has become important. While mixed state \gp was first introduced mathematically by Uhlmann,
recently Sj\"{o}qvist {\it et al.} [Phys. Rev. Lett. $\bf{85(14)}$, 2845 (2000)] have described the mixed state \gp in the context 
of quantum 
interference and shown theoretically that the visibility as well as the shift of the interference pattern are
functions of \gp and the purity of the mixed state. Here we report the first experimental study of the dependence 
of interference
visibility and shift of the interference pattern on the mixed state \gp by Nuclear Magnetic Resonance.  
\end{abstract}
\maketitle

\section{I. Introduction}
When a quantum system undergoes a unitary evolution and comes back to its initial state it acquires a 
phase. The acquired phase can be of two types; the dynamic phase which
depends on the evolution Hamiltonian and the geometric phase, which depends only on the evolution
path of the quantum system in the projective Hilbert space \cite{panch,berry,aharonov}. For a two level quantum system 
(spin-$\frac{1}{2}$), the projective Hilbert space is a
sphere and the geometric phase depends on the geodesical solid angle subtended at the 
center of the sphere by the path of evolution of the state vector. The concept of geometric phase first came in the adiabatic
context \cite{berry} but later Aharonov {\it et al.} \cite{aharonov} gave a non-adiabatic generalization of the theory of 
\gp. In the adiabatic approach the state vector is parallel
transported adiabatically to ensure that the system always remains in one of the eigenstates (assuming that the system initially is prepared in one of the
eigenstates) of the instantaneous Hamiltonian during the evolution. In the non-adiabatic approach the system is changed abruptly and the system comes back
to its initial state through different intermediate states. 
From the total phase acquired by the quantum system, the dynamical phase is eliminated by various methods in order to experimentally measure the geometric
phase. In magnetic resonance experiments this is achieved
by a spin echo \cite{ernstbook}. The theory of \gp of a  pure quantum system or pure state geometric phase is well understood and has been
demonstrated experimentally by various experimental systems such as Nuclear Magnetic Resonance (NMR) \cite{suter}, single \cite{kwiat} and 
two photon interferometry \cite{brendel}.\\\\
Recently it has been proposed that fault tolerant quantum computation can be performed using geometric
phase as it depends on the path and not on the speed of the evolution \cite{fault1,fault2}. To perform computation using geometric phase, it is necessary 
to understand the relation between geometric
phase and decoherence. As decoherence or relaxation leads a system from a pure state to a mixed state, an understanding of the mixed state 
geometric phase is needed. In 1986 Uhlmann mathematically introduced the concept of mixed state geometric phase \cite{uhlmann}. 
In this paper Uhlmann has taken a large system in pure state and a part or a subsystem
in mixed state and pointed out the unitary evolution in which the subsystem is transported in a maximally parallel
manner \cite{uhlmann}. Recently Sj\"{o}qvist {\it et al.} have provided a new description for mixed state \gp 
in terms of quantum interferometry \cite{sjoqvist}. In a quantum interferometer a quantum system undergoes a series of unitary evolutions,
after which the probability of finding the system in one of its eigenstates becomes an oscillatory function of some 
control parameter. The oscillation pattern of the probability resembles the well known optical interference pattern. 
According to Sj\"{o}qvist {\it et al.} the shift of interference pattern is a function of the 
\gp acquired by the quantum system during the unitary evolutions, as well as the purity of the initial internal state (such as, polarization of a photon) of 
the quantum systems involved in the
interferometric operation \cite{sjoqvist}. The geometric phase can be directly measured from the shift of the interference pattern. 
Mixed state geometric phase has been experimentally demonstrated by Du {\it et al.} \cite{du} 
using NMR and by Ericsson {\it et al.} \cite{eric} using single photon interferometry. Du {\it et al.} have experimentally 
demonstrated the mixed state geometric phase by measuring the relative phase change 
of an auxiliary spin. In the present work we measure the shift of the interference pattern in the Sj\"{o}qvist's interferometry model and show that the
shift is same as the theoretically predicted \gp as a function of mixed state purity. We also demonstrate the effect of mixed state geometric phase on the 
interference visibility. 
\section{II. Theory}
\subsection{Quantum Interference}
\begin{center}
\begin{figure}
\vspace*{-1cm}
\epsfig{file=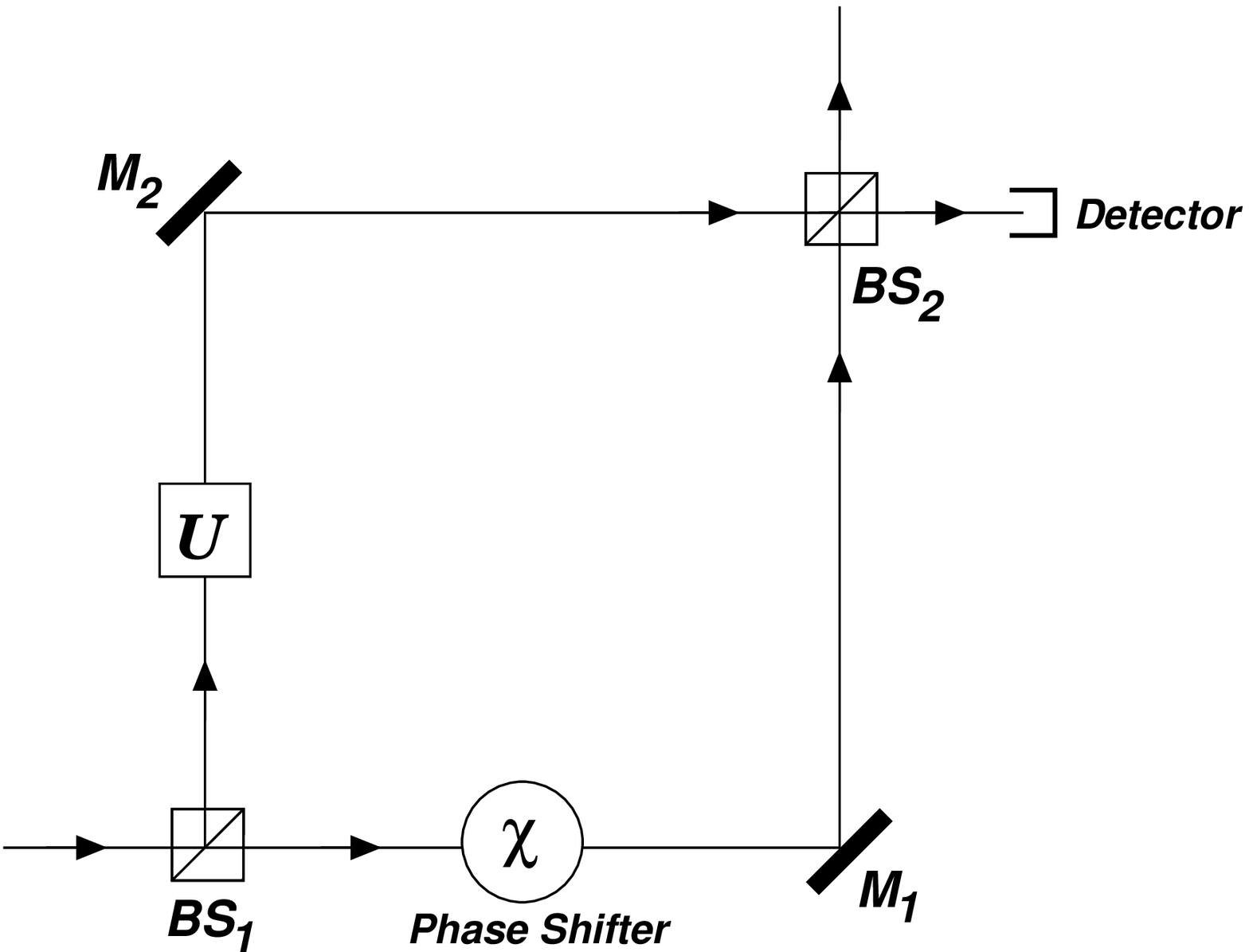,angle=0,width=0.8\textwidth}
\caption{}
\label{M-Z}
\end{figure}
\end{center}

\begin{center}
\begin{figure}
\vspace*{-1cm}
\epsfig{file=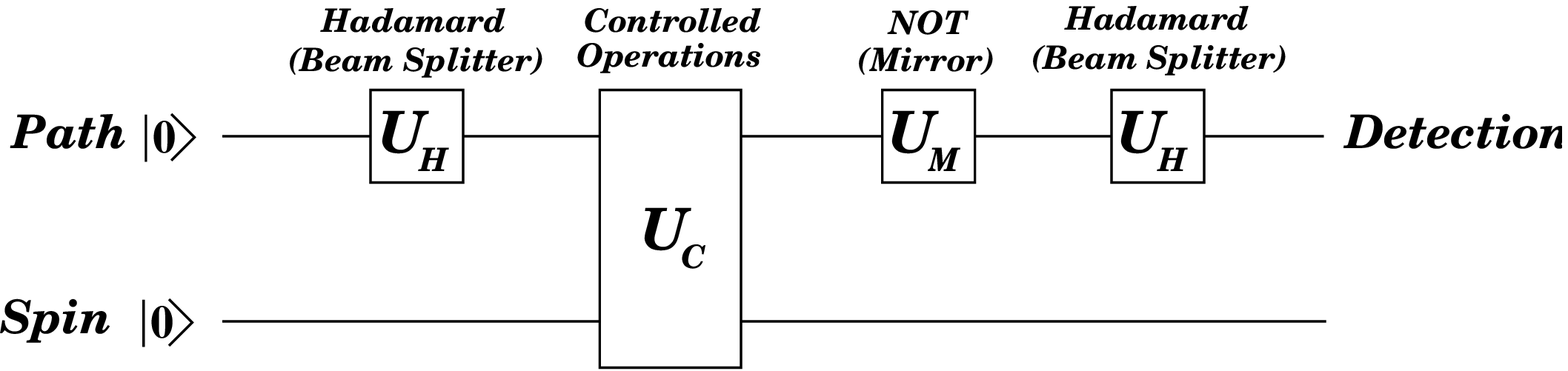,angle=0,width=0.8\textwidth}
\caption{}
\label{circuit}
\end{figure}
\end{center}

Let us consider the Sj\"{o}qvist's interferometry model as shown in Fig.\ref{M-Z} \cite{hosoya}. Photons entering the interferometer along a
horizontal path are split into two perpendicular paths by a beam splitter ($BS_1$). 
On the horizontal path the photons are
globally phase shifted, whereas on the other path the internal states of the photons, say the
polarization, undergo a unitary evolution {\it U}. The photons are reflected by two mirrors ($M_1$,$M_2$) and the two 
paths meet again at another beam splitter ($BS_2$). A detector detects the photons coming only along the
horizontal path. The detected intensity shows an interference pattern as a function of the phase shift.\\
As a photon can take one of the two possible paths and in each path it can have one of the two possible polarizations, so
the Hilbert space of the combined ``path-internal state" system becomes $2^{2}\times 2^{2}$. In NMR, the above 
interferometry
model can be simulated using two coupled spin-$\frac{1}{2}$ nuclei, which have the Hilbert space of identical
dimension. One qubit represents the path qubit and the other qubit, termed as spin qubit, represents the
internal state.\\\\
The equivalent quantum circuit of the Sj\"{o}qvist's interferometry model is shown in Fig.\ref{circuit} \cite{hosoya}. 
The eigenstates $|0\rangle$ and $|1\rangle$ of the path
qubit represent the two paths, horizontal and vertical respectively. The path qubit is prepared in the pure state
$|0\rangle \langle 0|$ at the beginning of the interferometry operation. The beam splitter is represented by a Hadamard gate
given by,
\begin{eqnarray}
U_H
=
\frac{1}{\sqrt{2}}
\left[
\begin{array}{cr}
1& 1\\
1&-1
\end{array}
\right].
\label{had}
\end{eqnarray}
As the phase shifter and the unitary operator {\it `U'} are path specific, they are represented by two 
controlled operations, together given by \cite{sjoqvist},
\begin{eqnarray}
U_C
=
\left[
\begin{array}{lr}
0&0\\
0&1
\end{array}
\right]
\otimes
U
+
\left[
\begin{array}{lr}
e^{i \chi}&0\\
0&0
\end{array}
\right]
\otimes
\mathds{1}.
\label{com}
\end{eqnarray}
Depending upon the state of the first qubit, the operator $U_C$ either applies {\it U} on
the second qubit or phase shifts the first qubit. The mirrors in Fig.\ref{M-Z} represent NOT gates, given by, 
\begin{eqnarray}
U_M
=
\left[
\begin{array}{lr}
0&1\\
1&0
\end{array}
\right].
\label{not}
\end{eqnarray}
The input state of the combined ``path-spin" system can be written as \cite{hosoya},
\begin{eqnarray}
\rho_{in} = (|0\rangle \langle 0|)^P \otimes \rho_0^S ,
\end{eqnarray}
where P stands for path and $\rho_0^S$ is the density matrix corresponding to the initial state of the spin 
qubit which can be either pure or mixed. The initial density matrix $\rho_{in}$ is transformed into the final density 
matrix $\rho_{out}$ as,
\begin{eqnarray}
\rho_{out} = U_H U_M U_C U_H \hspace*{0.1cm} \rho_{in} \hspace*{0.1 cm} {U_H}^{\dagger} {U_C}^{\dagger} {U_M}^{\dagger} 
{U_H}^{\dagger}.
\end{eqnarray}
Substituting the matrix forms of the operators given by Eq.(\ref{had}-\ref{not}) we obtain \cite{sjoqvist},
\small
\begin{eqnarray}
\rho_{out} = 
\frac{1}{4}
\left[
\left(
\begin{array}{lr}
1&1\\
1&1
\end{array}
\right)
\otimes
U\rho_0^S U^{\dagger}
+
\left(
\begin{array}{rr}
1&-1\\
-1&1
\end{array}
\right)
\otimes
\rho_0^S  
+
e^{i \chi}
\left(
\begin{array}{rr}
1&1\\
-1&-1
\end{array}
\right)
\otimes
\rho_0^S U^{\dagger}
+
e^{-i \chi}
\left(
\begin{array}{lr}
1&-1\\
1&-1
\end{array}
\right)
\otimes
U\rho_0^S
\right].
\end{eqnarray}
\large  
The detected signal in the horizontal path ($|0\rangle$ eigenstate of path qubit) is given by the trace of the reduced
density matrix of the spin qubit corresponding to the $|0\rangle$ state of the path qubit. The output intensity is given
by \cite{hosoya},
\be
I &=& \frac{1}{4} Tr_S \left(U\rho_0^S U^{\dagger} + \rho_0^S + e^{-i \chi}U\rho_0^S + e^{i \chi}\rho_0^S U^{\dagger} \right)
\nonumber\\
&=& \frac{1}{2}\left(1 + |Tr_S \left(U\rho_0^S \right)|cos\left[\chi - arg \  Tr_S \left(U\rho_0^S \right)\right]\right) \nonumber
\\ 
&=& \frac{1}{2}\left(1 + \nu \ cos\left[\chi - \phi \right]\right),
\label{intpat}
\ee 
where the amplitude of oscillation $\nu = |Tr_S \left(U\rho_0^S \right)|$ is called the visibility of interference and the 
shift $\phi = arg  \ Tr_S \left(U\rho_0^S \right)$ depends on the unitary operator {\it `U'} acting on the spin qubit density 
matrix $\rho_0^S$.\\\\
A mixed state can be thought of a mixture of several pure states incoherently weighted by their respective
probabilities. Therefore, the interference pattern given by Eq.\ref{intpat} takes the following form for a mixed input
spin state\cite{sjoqvist},
\be
I = \sum_k w_k I_k = \frac{1}{2} \left( 1 + \sum_k w_k \nu_k \ cos\left[ \chi - \phi_k \right] \right),
\label{intmix}
\ee       
where the index k denotes individual pure states with probabilities $w_k$. The above equation can be written in the form of Eq.\ref{intpat} as
\be
I \propto 1 + \tilde{\nu}\ cos(\chi - \tilde{\phi}),
\ee
by defining mixed state phase shift $\tilde{\phi}$ and visibility $\tilde{\nu}$ as\cite{sjoqvist},
\be
\tilde{\phi}& = & arg \ \left(\sum_k w_k \nu_k e^{i \phi_k} \right),
\label{mixphi} 
\ee
\be
\tilde{\nu}& =& \left| \sum_k w_k \nu_k e^{i \phi_k} \right|.
\label{mixnu}
\ee  
\subsection{Geometric phase and parallel transport condition}
The parallel transport condition for any state vector $|\psi (t)\rangle$ is given by,
\be
\langle \psi (t)|\dot{\psi}(t)\rangle = 0,
\label{parcon}
\ee
which means that the phase does not change when $|\psi (t)\rangle$ evolves to $|\psi (t+\delta t)\rangle$ for infinitesimal
$\delta t$. When a mixed
state given by the density matrix $\rho_m (t)$ evolves under a unitary operator {\it A(t)}, the condition given in
Eq.\ref{parcon} leads to \cite{sjoqvist},
\be
Tr \left[ \rho_m (t) \dot{A}(t) A^{\dagger}(t)\right] = 0.
\ee
This condition although necessary is not sufficient to determine the unitary operator {\it A(t)} as it can determine {\it
A(t)} only up to N phase factors, where N is the dimension of the Hilbert space. The N phase factors can be determined from the
conditions \cite{wagh},
\be
\langle k(t)\left| \dot{A}(t)\ A^{\dagger}(t)\right|k(t)\rangle = 0, \hspace*{2cm} k = 1,2,3......,N,
\label{strcon}
\ee
where $|k(t)\rangle$ are the orthonormal eigenstates of $\rho_m (t)$. The unitary operator {\it A(t)}, obtained by
solving the above conditions, parallel transports the mixed state density matrix $\rho_m$ so that the dynamical phase
becomes identically zero.  
The \gp $\gamma_g$, acquired by a mixed state when the state evolves under {\it A(t)} along a curve $\Gamma$, is given by \cite{sjoqvist},
\be
\gamma_g[\Gamma] = arg\ Tr[\rho_m A(t)] = arg\left( \sum_k w_k \nu_k e^{i \beta_k}\right),
\label{mgp}
\ee
where $e^{i \beta_k}$ is the \gp associated with the $k^{th}$ pure state. The 
expression for the \gp given by Eq.\ref{mgp} is similar to the
expression for the interferometric phase shift given by Eq.\ref{mixphi} and therefore the interferometric phase shift 
directly gives the \gp of the spin qubit.\\\\
In the present work we consider the mixed state of a spin-$\frac{1}{2}$ particle. The density operator of a spin-$\frac{1}{2}$ particle can be in 
general written as,
\be
\rho_m = \frac{1}{2} \left( 1 + \vec{r}.\vec{\sigma} \right),
\ee
where the length `r' of the Bloch vector $\vec{r}$, is equal to one for pure states, less than one for mixed states and 
remains unchanged during unitary evolution of the state. The components of $\vec{\sigma}$ are the Pauli matrices,
$\vec{\sigma} = [\sigma_x,\sigma_y,\sigma_z]$. $\rho_m$ represents a mixture of two of its eigenvectors with eigenvalues
$\frac{1}{2}\left( 1 \pm r\right)$.\\
Let us consider that the Bloch vector $\vec{r}$ for a mixed state (r$<$1) traces out a cyclic curve in the Bloch sphere which 
subtends a geodesically closed solid angle of $\Omega$. During the process the two eigenstates of the density operator with eigenvalues 
$\frac{1}{2}\left( 1 \pm r\right)$ acquire \gp $\mp \frac{\Omega}{2}$ respectively\cite{anandan}. The quantity $\sum_k w_k
\nu_k e^{i \phi_k}$ (Eq.\ref{mixphi} and \ref{mixnu}), using the fact that $\nu_k = 1$ for cyclic evolution\cite{sjoqvist}, becomes,
\be
\sum_k w_k \nu_k e^{i \phi_k}& =& \frac{1}{2}\left(1-r \right)e^{i \frac{\Omega}{2}} + \frac{1}{2}\left(1+r \right)e^{-i
\frac{\Omega}{2}}\nonumber \\
&=& cos(\frac{\Omega}{2}) - i\  r\  sin(\frac{\Omega}{2}).
\label{quan}
\ee      
Using the expression given by Eq.\ref{quan} 
the shift of interference pattern (Eq.\ref{mixphi}) and the visibility (Eq.\ref{mixnu}) for mixed state can be respectively written as,
\be
\tilde{\phi} = -\ arctan\left( r\  tan \left(\frac{\Omega}{2}\right)\right),
\label{shiftgeo}
\ee
and
\be
\tilde{\nu} = \sqrt{cos^{2}(\frac{\Omega}{2}) + r^{2} sin^{2}(\frac{\Omega}{2})}.
\label{visgeo}
\ee
In the present paper, we have experimentally measured the above frequency shift and the visibility using a two qubit system,
by NMR. The frequency shift directly gives the geometric phase of the spin qubit.
\section{III. Experimental Procedure}
\begin{center}
\begin{figure}
\vspace*{-1cm}
\epsfig{file=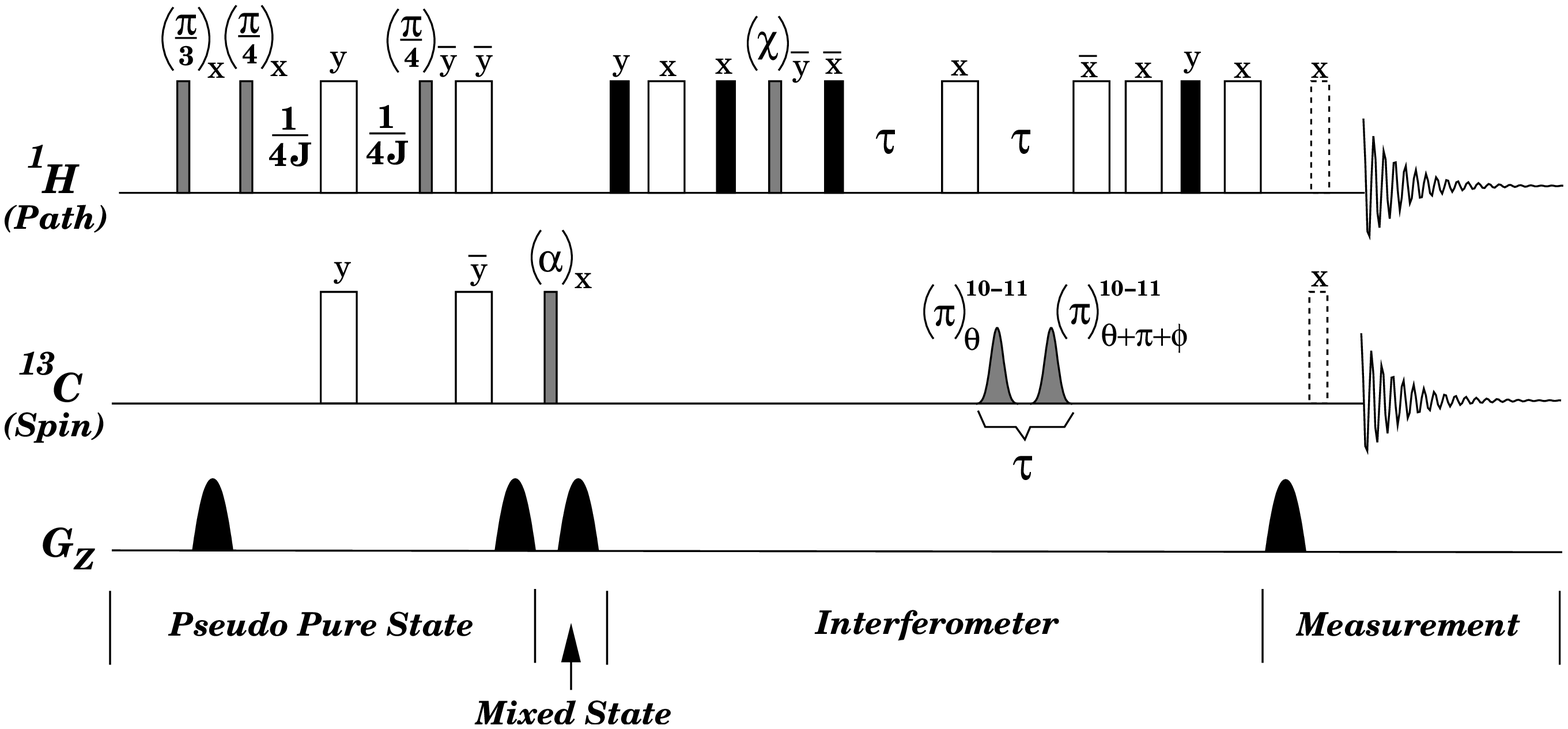,angle=0,width=0.9\textwidth}
\caption{}
\label{pp}
\end{figure}
\end{center}

Experiments were performed on Carbon-13 enriched $^{13}CHCl_3$ dissolved in $CDCl_3$. The $^{13}C$ and $^{1}H$ nuclei form a two 
qubit system with a J-coupling of 209 Hz. $^{1}H$ and $^{13}C$ respectively are used as the path and spin qubits. 
The spin-lattice ($T_1$) relaxation times of $^{13}C$ and $^{1}H$ at room temperature were measured
as 21s and 16s respectively, and the spin-spin ($T_2$) relaxation times were measured to be 0.29s and 3.4s respectively. All the experiments were 
performed using a 
Bruker DRX 500 MHz (11.2 Tesla) NMR spectrometer where the resonance frequencies for $^{13}C$ and $^{1}H$ are 125.76 MHz and 500.13 MHz 
respectively. The pulse
programme is given in Fig.\ref{pp}. The pulse programme contains several parts which are described below:
\subsection{Creation of Pseudo-pure state (PPS)}
The ``path-spin" system is first prepared in a pseudo-pure state
using the method of spatial averaging \cite{cory98}. The pulse sequence is as follows,
\be
\left(\frac{\pi}{3}\right)_x^H - G_z - \left(\frac{\pi}{4}\right)_x^H - \frac{1}{4J_{CH}} - \left(\pi \right)_y^{H,C} - \frac{1}{4J_{CH}}
-\left(\frac{\pi}{4}\right)_{\bar{y}}^H - \left(\pi\right)_{\bar{y}}^{H,C} - G_z
,
\ee
where the superscript H or C identifies the spin (proton or carbon respectively) on which the r.f. pulse is applied and the subscript {\it x} or {\it y} 
determines the phase of the pulse. $J_{CH}$ is
the J-coupling and $G_z$ indicates a {\it z}-gradient which destroys all coherences ({\it x} and {\it y} magnetizations) and retains only longitudinal 
magnetization ({\it z} magnetization component). At the end of this sequence the system is prepared in the $|00\rangle$ pseudo pure statei \cite{cory98}.  
\subsection{Creation of Mixed state}
After preparing the $|00\rangle$ PPS, an $\alpha$ degree pulse is applied on the carbon spin followed by a {\it z}-gradient. 
In the Bloch representation it creates a mixed state vector 
whose 
length r = $cos\ \alpha$ $[r < 1$ for $0^{\circ} < \alpha \leq 90^{\circ}$], where the value of r determines the purity of the state. 
The above pulse
programme can be written as,
\be
\left( \alpha \right)_x^C - Gz.
\ee    
\subsection{The Interferometer}
\begin{center}
\begin{figure}
\vspace*{-1cm}
\epsfig{file=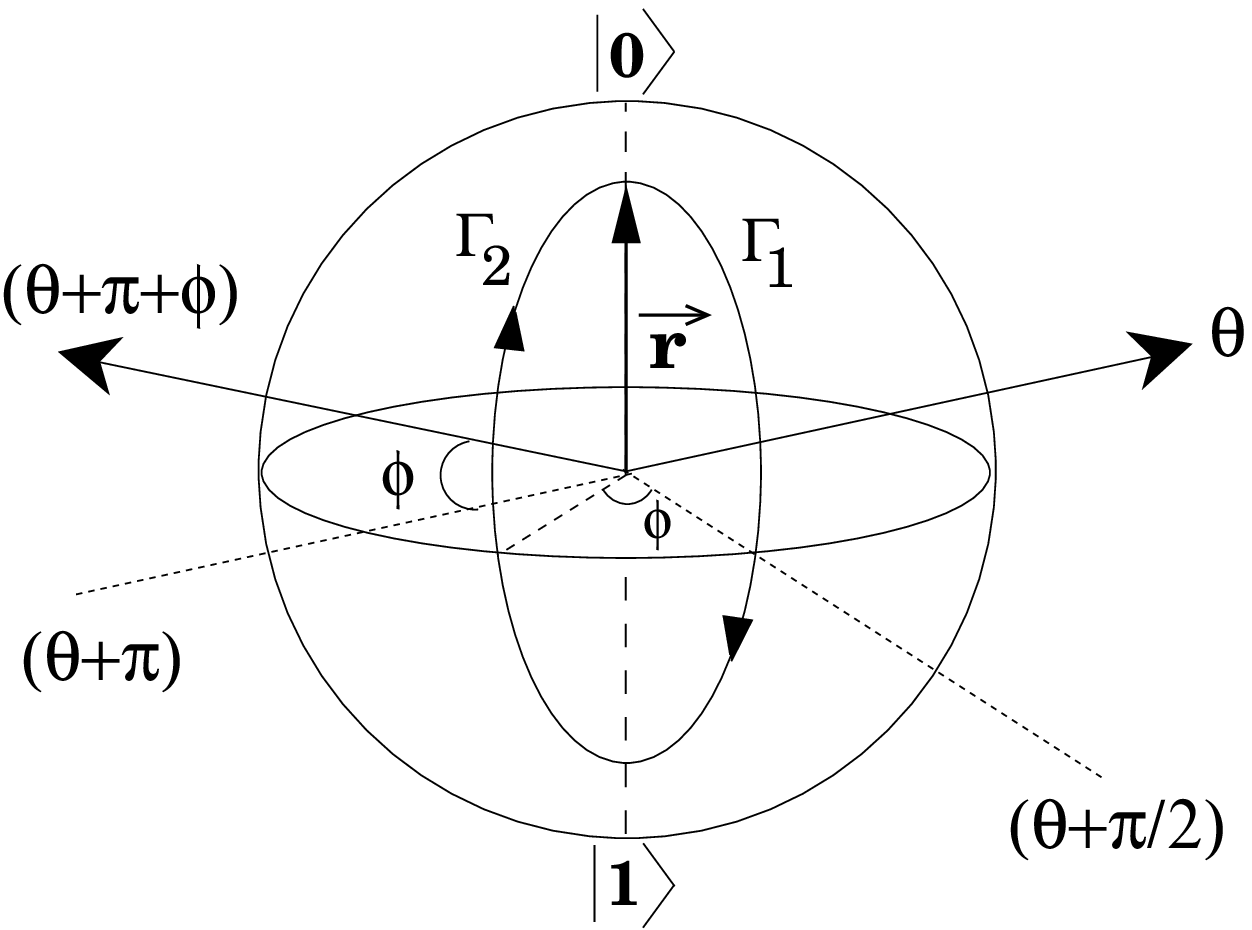,angle=0,width=0.8\textwidth}
\caption{}
\label{cycle}
\end{figure}
\end{center}

The Hadamard gate (Beam Splitter) is implemented by the sequence $\left(\frac{\pi}{2}\right)_y^H -
\left(\pi\right)_x^H$ \cite{jonesgate}. The important operation of the 
interferometry part is the controlled operation $U_C$. $U_C$ contains a controlled phase shift gate applied on the path 
qubit and a
controlled {\it U} operation acting on the spin qubit.\\
\small {\bf Controlled Phase shift:} \large\\
\hspace*{0.5cm}  
The controlled phase shift gate in the present context is different from the conventional two-qubit gate. Here both the controlling and the target qubits 
are the path qubit. The path qubit is phase
shifted by $\chi$ when it is in the $|0\rangle$ (horizontal path) state. The output of the phase shift gate $U_{\chi}$ can
be written as,
\be
U_\chi|00\rangle = e^{i \chi}|00\rangle ;\:\nonumber 
U_\chi|01\rangle = e^{i \chi}|01\rangle ;\:\nonumber
U_\chi|10\rangle = |10\rangle ;\:\nonumber 
U_\chi|11\rangle = |11\rangle.
\ee 
The pulse sequence for $U_\chi$ is \cite{jonesgate},
\be
\left(\frac{\pi}{2}\right)_x^H - \left( \chi \right)_{\bar{y}}^H - \left(\frac{\pi}{2}\right)_{\bar{x}}^H
\ee
\small {\bf Controlled {\it U}:} \large\\
\hspace*{0.5cm}  
In the present case {\it U} is a geometric phase shift operator. The controlled geometric phase shift operator is
implemented by evolving the spin qubit in a cyclic path (the `Slice Circuit') in the Bloch sphere as shown in Fig.\ref{cycle}, when the path qubit is in
state $|1\rangle$ (vertical path). This is achieved by pulsing only on the $|10\rangle - |11\rangle$ 
subsystem. Two transition selective $\pi$
pulses are applied on the $|10\rangle - |11\rangle$ transition of $^{13}C$ with phases differing by $(\pi + \phi)$ \cite{ranathesis}. They 
cause the Bloch vector to flip along one path, $\Gamma_1$ and come back to its initial orientation
along a different path, $\Gamma_2$. The loop ($\Gamma_1,\Gamma_2$) subtends a geodesically closed solid angle of 
$\Omega = 2\phi$ \cite{suter}.
A spin echo sequence has been incorporated to eliminate the dynamical phase.
The pulse sequence is given by,
\be
\tau - \left(\pi \right)_x^H - \left(\pi\right)_{\theta}^{|10\rangle - |11\rangle} -
\left(\pi\right)_{\theta + \pi + \phi}^{|10\rangle - |11\rangle} - \left(\pi \right)_{\bar{x}}^H,
\ee
where $\tau$ is the total time of the two transition selective pulses. The second $\left(\pi \right)^H_{\bar{x}}$ pulse 
in the above sequence restores the sign of the $^{1}H$ magnetization.\\\\
The mirror (NOT gate) is implemented by a $\left(\pi\right)_x^H$ pulse. It converts the state $|0\rangle$ (horizontal path) to state $|1\rangle$ 
(vertical path) and vice-versa. The sequence $\left(\frac{\pi}{2}\right)_y^H - \left(\pi\right)_x^H$ for the Hadamard gate 
is repeated after the mirror, in order to implement the second beam splitter ($BS_2$).   

\subsection{Measurement}
At the end of the interferometric operations, both the 
qubits were detected after applying a {\it z}-gradient and a reading $\pi/2$ pulse on the detection qubit. The diagonal
part of the density matrix was then tomographed using the line intensities normalized to the respective equilibrium
spectra \cite{rana}.
\section{IV. Results} 
\begin{center}
\begin{figure}
\vspace*{-1cm}
\epsfig{file=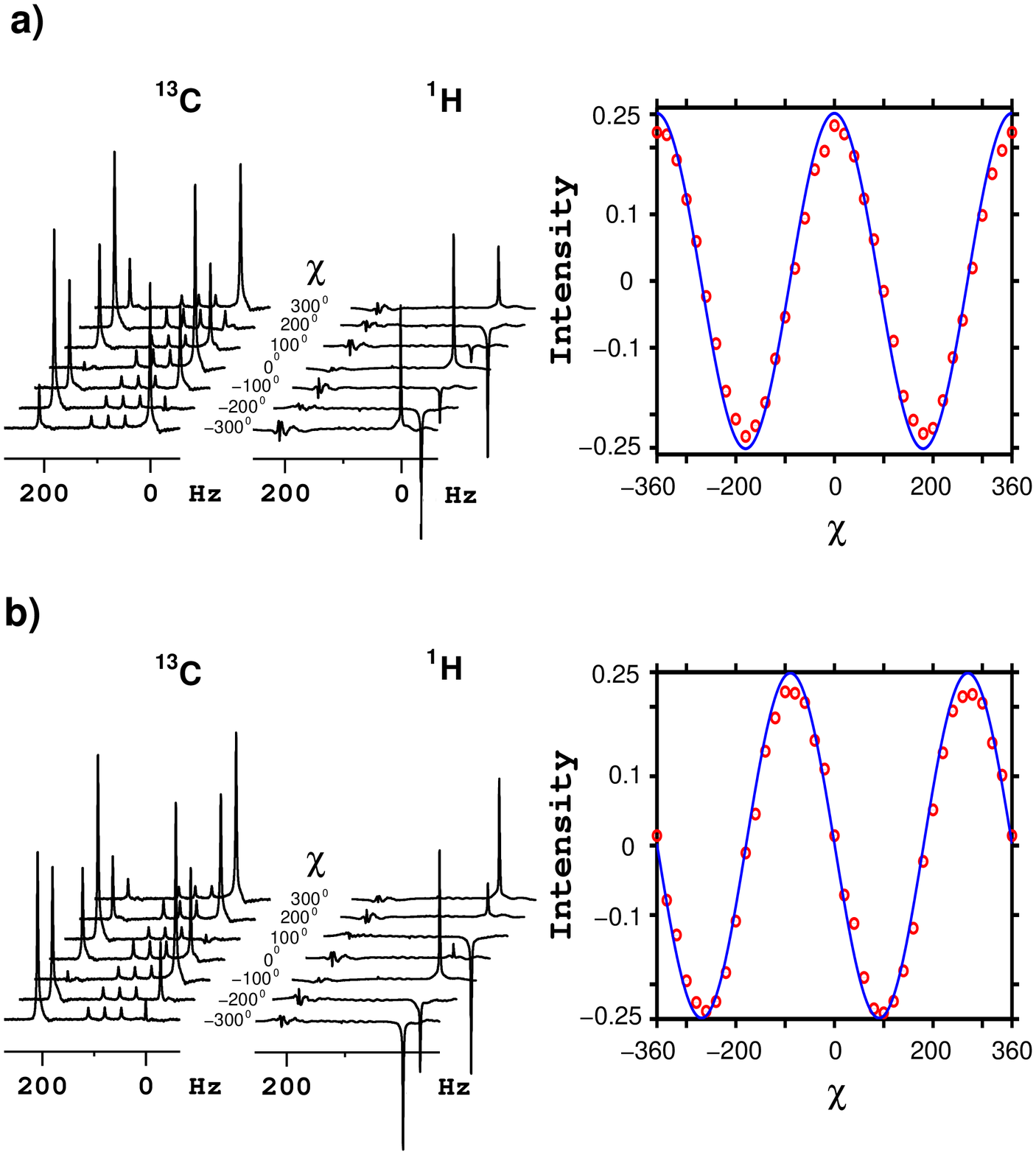,angle=0,width=0.9\textwidth}
\caption{}
\label{pattern}
\end{figure}
\end{center}

The intensity of the signal detected only in the horizontal path is proportional to the total population of the
$|00\rangle$ and
$|01\rangle$ levels as these two energy levels correspond to the $|0\rangle$ state (horizontal path) of the path qubit. For
each $\chi$ the final density matrix (diagonal part only) was tomographed and the sum of the $|00\rangle$ and 
$|01\rangle$ populations was plotted as the intensity.
Data were collected at 37 equidistant values of $\chi$ ranging from -360$^\circ$ to 360$^\circ$ to obtain the full 
interference pattern. 
Fig.\ref{pattern} shows the $^1H$ and $^{13}C$ spectra for different $\chi$ values for pure initial state
of the spin qubit. The three low intensity lines in the carbon spectra arise from the natural abundant carbon coupled to deuterium in the solvent
CDCl$_3$. The intensities (sum of $|00\rangle$ and $|01\rangle$ populations) calculated from the normalized 
spectral line intensities are plotted
as a function of phase shift $\chi$. Fig.\ref{pattern}(a) shows the pattern when {\it U} (the geometric phase shift 
operator) was not applied.
As expected no shift in the pattern from $\chi$ = 0 was observed. Whereas Fig.\ref{pattern}(b) shows the pattern
corresponding to $\Omega$ = 180$^\circ$. A shift of -90$^\circ$ was observed as expected according to Eq.\ref{shiftgeo} for
pure state (r = 1). In each plot the solid line represents the expected theoretical curve.         

\subsection{The `shift - geometric phase' relationship (Eq.\ref{shiftgeo})}
\begin{center}
\begin{figure}
\vspace*{-1cm}
\epsfig{file=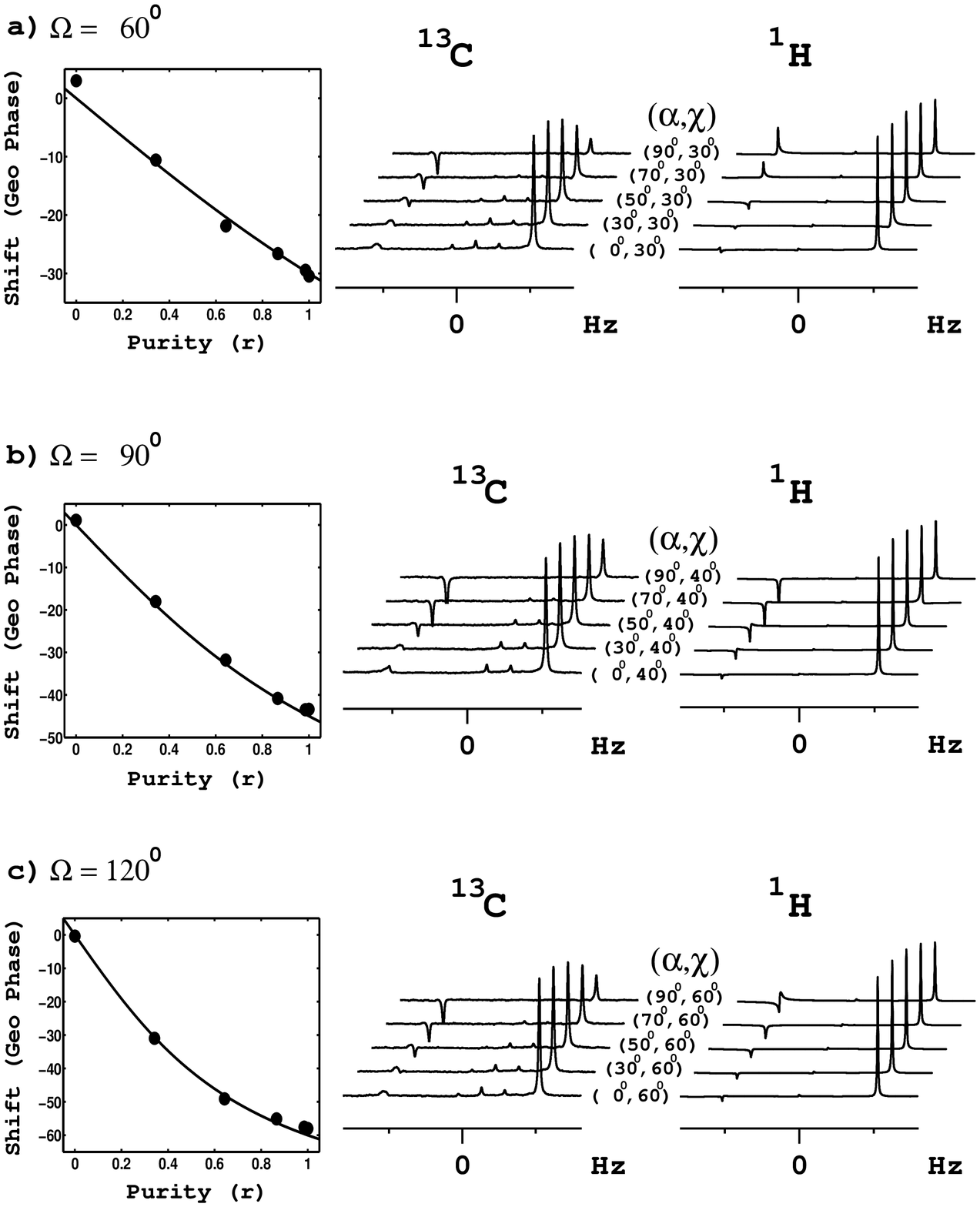,angle=0,width=0.9\textwidth}
\caption{}
\label{shift}
\end{figure}
\end{center}

For mixed input state of the spin qubit, the pattern shifts from $\chi$ = 0 for non-zero geometric phase. The amount of 
shift is a
function of both the purity of the input state as well as the \gp of the spin qubit. Fig.\ref{shift} shows the dependence
of interferometric shift on the \gp and the purity of mixed state for $\Omega$ = 60$^\circ$ (\ref{shift}.a), 
$\Omega$ = 90$^\circ$ (\ref{shift}.b) and 
$\Omega$ = 120$^\circ$ (\ref{shift}.c). For a particular value of $\Omega$ and r, experiments have been performed for ten 
equidistant points of $\chi$ in the range [-90$^\circ$,0$^\circ$]. The data were fitted with a function 
$\mathscr{F}$($\nu,\phi$) = $\nu \:$cos($\chi - \phi$), to calculate the shift. The shift is zero for r = 0 and 
the shift is $-\frac{\Omega}{2}$ for r = 1 in all 
the three cases. The solid line in each plot represents the theoretical
curve. Spectra corresponding to $\alpha$ = 0$^\circ$,30$^\circ$,50$^\circ$,70$^\circ$ and 90$^\circ$ have been shown for
$\chi$ = 30$^\circ$ (\ref{shift}.a), $\chi$ = 40$^\circ$ (\ref{shift}.b) and $\chi$ = 60$^\circ$ (\ref{shift}.c). 
\subsection{The `visibility - geometric phase' relationship (Eq.\ref{visgeo})}
\begin{center}
\begin{figure}
\vspace*{-1cm}
\epsfig{file=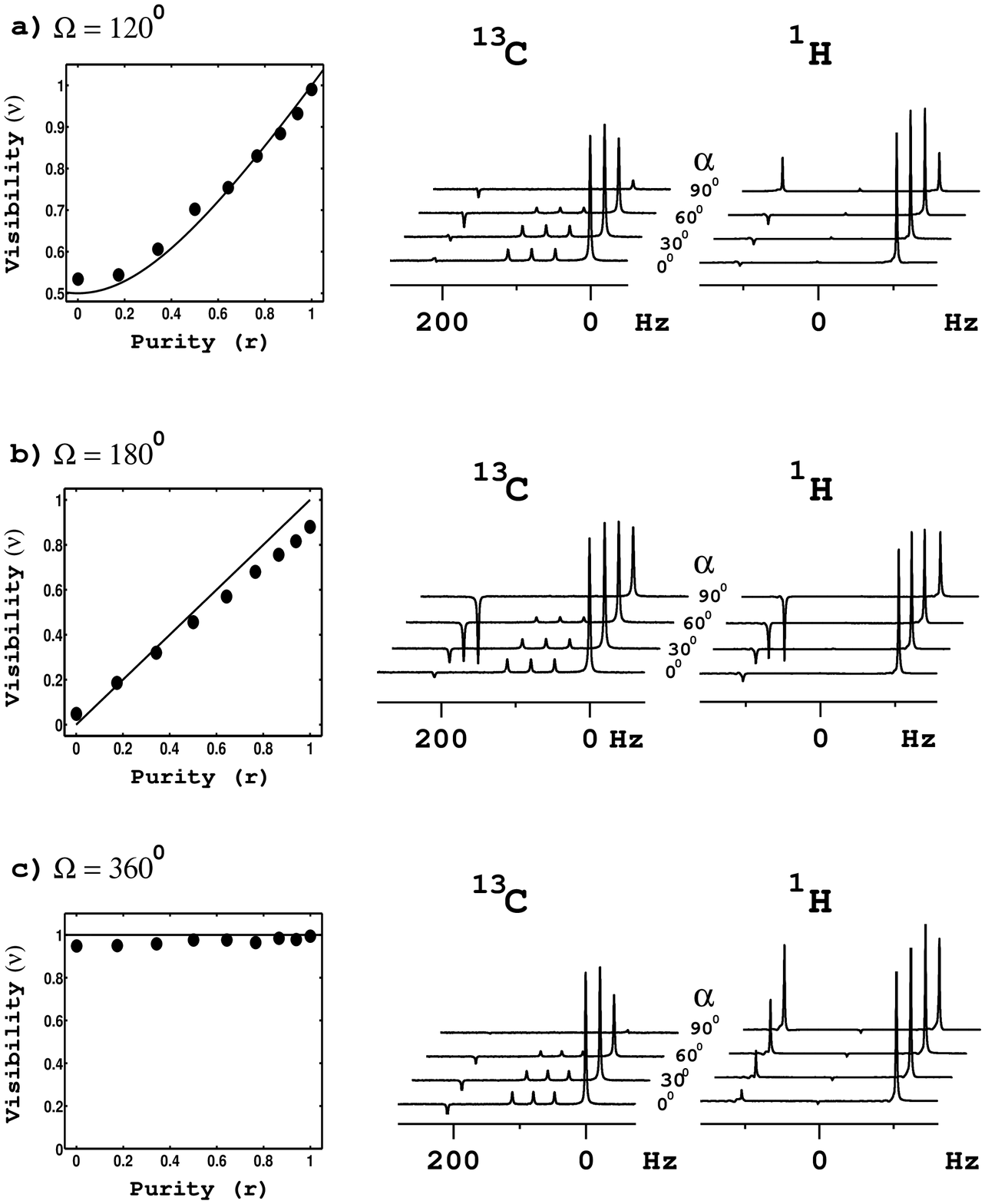,angle=0,width=0.9\textwidth}
\caption{}
\label{visib}
\end{figure}
\end{center}

The visibility
of interference or the amplitude of oscillation is given by the difference between the maximum and the minimum intensities
in the interference pattern. The visibility was measured for different purity `r' of the spin qubit state.
Fig.\ref{visib} shows the visibility as a function of `r' for $\Omega$ = 120$^\circ$ (\ref{visib}.a), $\Omega$ = 180$^\circ$
 (\ref{visib}.b) and $\Omega$ = 360$^\circ$ (\ref{visib}.c). $\Omega$ = 360$^\circ$ makes the visibility independent of r, 
while for $\Omega$ = 180$^\circ$ 
the visibility changes linearly with r. The experimental data matches the expected behavior (solid line) given by 
Eq.\ref{visgeo}. Spectra corresponding to
$\alpha$ = 0$^\circ$,30$^\circ$,60$^\circ$ and 90$^\circ$ (for $\Omega$ = 180$^\circ$, $\alpha$ = 89$^\circ$ was applied
instead of 90$^\circ$ as the shift according to Eq.\ref{shiftgeo} becomes undefined for $\Omega$ = 180$^\circ$ and r = 0) have 
been shown adjacent to each plot. While recording the spectra
the value of $\chi$ was chosen according to the shift of pattern given by Eq.\ref{shiftgeo}. All the experiments for 
$\Omega$ = 360$^\circ$ was performed at $\chi$ = 0$^\circ$ and for $\Omega$ = 180$^\circ$, at $\chi$ = -90$^\circ$. For
$\Omega$ = 120$^\circ$, $\chi$ was chosen same as the shift predicted by Eq.\ref{shiftgeo}.     
 
\section{V. Conclusion}

The study of mixed state geometric phase has become important ever since \gp was proposed as a possible method of
performing fault tolerant quantum computing. The pure state \gp is well understood and well studied by various
experimental methods. Here we have reported the first 
experimental measurement of mixed state \gp directly from the shift of a quantum interference pattern. We have 
experimentally measured the visibility and the shift of the interference pattern as a function of the purity of the input 
mixed state which agree with the theoretically expected results. This study shows that NMR interferometry is one of the 
possible experimental methods to measure \gp of a pure as well as a mixed state. Future directions include studies of 
non-cyclic \gp \cite{sjo-pla} and applications of \gp in fault tolerant quantum computations. 

\section*{Acknowledgments}
We gratefully acknowledge Prof. K. V. Ramanathan for discussions. The use of DRX-500 high resolution liquid state
spectrometer of the NMR Research Centre, Indian Institute of
Science, Bangalore, funded by Department of Science and Technology (DST), New Delhi, is gratefully acknowledged. AK
acknowledges ``DAE-BRNS" for
``Senior Scientist scheme", and DST for a research grant.
\newpage

\newpage
\begin{center}
{\bf FIGURE CAPTIONS}
\end{center}
Figure 1. The Sj\"{o}qvist's interferometry model \cite{hosoya}. BS$_1$ and BS$_2$ are two beam splitters which split a photon beam in two
perpendicular paths. M$_1$ and M$_2$ are two mirrors. The unitary operation {\it `U'} acts on the internal state of the photons
traveling along the vertical path while the photons traveling along the horizontal path are phase shifted by $\chi$ by 
the phase shifter. The detector detects the photons coming along the horizontal path only.\\\\
Figure 2. The quantum equivalent circuit of Sj\"{o}qvist's interferometry model using two qubits \cite{hosoya}. 
One qubit corresponds the
path while the other corresponds the spin (internal state). The single qubit gates used are Hadamard ($U_H$, Eq.\ref{had})
and NOT ($U_M$, Eq.\ref{not}). The control operation $U_C$ consists of a control phase shift gate acting on the first qubit
and a control {\it `U'} operator acting on the second qubit (Eq.\ref{com}).\\\\
Figure 3. The pulse programme to perform NMR interferometry. The black and empty boxes represent $\pi/2$ and $\pi$ pulses 
respectively while the gray boxes represent pulses with flip angle given on the top. The phase of a pulse is given at top 
of the pulse. The gray
Gaussian shaped pulses are transition selective soft $\pi$ pulses applied on the $|10\rangle - |11\rangle$ Carbon transition. 
G$_Z$ is the {\it z}-Gradient pulses applied to kill all transverse magnetizations. J is the coupling between $^{1}H$ and
$^{13}C$ and is equal to 209 Hz in the present case. The pulse programme can be divided in four major parts, namely the PPS, preparation of 
mixed state, 
interferometer and the measurement, which are discussed in detail in the text. The last $\pi/2$ pulses (shown by broken
lines) on each spin is the
detection pulse and is applied at the detected spin one at a time.\\\\
Figure 4. The slice circuit for cyclic evolution of a state vector [for pure state $|\vec{r}|$ = 1 and for mixed states $|\vec{r}|$ $<$ 1] in the 
Bloch sphere \cite{ranathesis}. 
In the present case, the $|10\rangle - |11\rangle$ subsystem is transported through a closed loop using two
transition selective $\pi$ pulses applied at phases ($\theta$) and ($\theta+\pi+\phi$) on the $|10\rangle - |11\rangle$ 
Carbon transition. The first $\pi$ pulse takes
the second qubit from state $|0\rangle$ to state $|1\rangle$ along the path $\Gamma_1$, while the second $\pi$ pulse brings
the state back from $|1\rangle$ to $|0\rangle$ along a different path $\Gamma_2$. The loop [$\Gamma_1$,$\Gamma_2$] subtends a solid angle
$\Omega$ = 2$\phi$ at the center of the Bloch sphere \cite{suter}.\\\\
Figure 5. The $^{1}H$ and $^{13}C$ spectra obtained for various values of $\chi$, starting from pseudo pure initial state of the spin qubit. 
The additional low intensity lines in the carbon spectra arise from the natural abundant carbon coupled to deuterium in the solvent CDCl$_3$.
The sum of $|00\rangle$ and $|01\rangle$ populations, calculated from the normalized line intensities have been plotted as
intensities against 37 equidistant values of the phase shift $\chi$ in the range [-360$^\circ$,360$^\circ$]. (a) $\Omega$ = 0$^\circ$. No shift in 
pattern was 
observed. (b) $\Omega$ = 180$^\circ$. A shift of -90$^\circ$ was observed. 
The solid line in each plot represents the theoretically expected value [Eq.\ref{intpat}].\\\\
Figure 6. The shift in interference pattern as a function of purity of mixed state for $\Omega$ = 60$^\circ$ (a),
$\Omega$ = 90$^\circ$ (b) and $\Omega$ = 120$^\circ$ (c). In each plot the shift is plotted for 6 different values of 
r (cos$\alpha$) between 0 and 1. $^{13}C$ and $^{1}H$ spectra are shown for 5 different pairs of $\alpha$ and $\chi$.
The solid line in each plot represents the theoretically expected curve according to Eq.\ref{shiftgeo}.\\\\
Figure 7. The dependence of interference visibility on mixed state purity for $\Omega$ = 120$^\circ$ (a),
$\Omega$ = 180$^\circ$ (b) and $\Omega$ = 360$^\circ$ (c). In each plot the visibility is 
plotted for 9 different values of r (cos$\alpha$) ranging from 0 to 1 (for $\Omega$ = 180$^\circ$, $\alpha$ was varied from
0$^\circ$ to 89$^\circ$ instead of 90$^\circ$ in order to avoid the singularity of Eq.\ref{shiftgeo}). $^{13}C$ and $^{1}H$ spectra 
are shown for 4 different values of $\alpha$. The solid line in each plot represents the theoretically expected curve 
according to Eq.\ref{visgeo}.
\newpage
\newpage
\newpage
\newpage
\newpage
\newpage
\newpage
\end{document}